\documentclass[sigconf,authorversion,9pt]{acmart}
\usepackage{amsmath}

\usepackage{amssymb,amsfonts}
\usepackage{algorithmic}
\usepackage{graphicx}

\usepackage{textcomp}
\usepackage{xcolor}
\def\BibTeX{{\rm B\kern-.05em{\sc i\kern-.025em b}\kern-.08em
    T\kern-.1667em\lower.7ex\hbox{E}\kern-.125emX}}
\usepackage[utf8]{inputenc}
\usepackage{booktabs} 
\usepackage{graphicx}
\usepackage{subfigure}
\usepackage{units}
\usepackage{blindtext}
\usepackage{multirow}
\usepackage{xcolor,xspace}
\usepackage{amsmath}
\usepackage{paralist}
\usepackage{mdframed}
\usepackage{pict2e}
\usepackage{mdframed}
\usepackage{listings}
\usepackage{color}
\usepackage{threeparttable}
\usepackage{array}
\usepackage{booktabs}
\usepackage{pifont}
\usepackage{balance}
\usepackage{url}
\usepackage{verbatim}

\usepackage{lipsum}

\usepackage[ruled]{algorithm2e}
\usepackage{algorithmic}
\LinesNumbered

\usepackage{url}

\usepackage[
    n,
    advantage,
    operators,
    sets,
    adversary,
    landau,
    probability,
    notions,
    logic,
    ff,
    mm,
    primitives,
    events,
    complexity,
    asymptotics,
    keys]{cryptocode}

\usetikzlibrary{arrows, patterns, automata}

\newcommand{\ignore}[1]{}
\newcommand{\edit}[1]{\textcolor{black}{#1}}

\newcommand{\acro}{{{\it $ASAP$}}\xspace}
\newcommand{\acrotextcapital}{{\textit{{\bf A}rchitecture for {\bf S}ecure {\bf A}synchronous {\bf P}rocessing in PoX}}\xspace}

\newcommand{\pox}{{{\sf\it PoX}}\xspace}
\newcommand{\vrased}{{{\sf\it VRASED}}\xspace}

\newcommand{\dev}{{\ensuremath{\sf{\mathcal Prv}}}\xspace}
\newcommand{\prv}{{\ensuremath{\sf{\mathcal Prv}}}\xspace}
\newcommand{\vrf}{{\ensuremath{\sf{\mathcal Vrf}}}\xspace}

\newcommand{\RA}{{\ensuremath{\sf{\mathcal RA}}}\xspace}

\newcommand{\chal}{{\ensuremath{\sf{\mathcal Chal}}}\xspace}

\newcommand{\attkey}{\ensuremath{\mathcal K}\xspace}

\mathchardef\mhyphen="2D

\newcommand{\hw}{\texttt{\small HW-Mod}\xspace}

\renewcommand\footnotetextcopyrightpermission[1]{} 
\begin{document}

\title[\acro: Reconciling Asynchronous Real-Time Operations and Proofs of Execution]
      {\acro: Reconciling Asynchronous Real-Time Operations and Proofs of Execution in Simple Embedded Systems}


\author{Adam Caulfield}
\affiliation{%
	\institution{Rochester Institute of Technology}
}

\author{Norrathep Rattanavipanon}
\affiliation{%
	\institution{Prince Songkla University}
}

\author{Ivan De Oliveira Nunes}
\affiliation{%
	\institution{Rochester Institute of Technology}
}

\begin{abstract}
Embedded devices are increasingly ubiquitous and their importance is hard to overestimate.
While they often support safety-critical functions (e.g., in medical devices and sensor-alarm combinations), they are usually implemented under strict cost/energy budgets, using low-end microcontroller units (MCUs) that lack sophisticated security mechanisms.
Motivated by this issue, recent work developed architectures capable of generating Proofs of Execution (PoX) for the correct/expected software in potentially compromised low-end MCUs.
In practice, this capability can be leveraged to provide ``integrity from birth'' to sensor data, by binding the sensed results/outputs to an unforgeable cryptographic proof of execution of the expected sensing process.
Despite this significant progress, current PoX schemes for low-end MCUs ignore the real-time needs of many applications.
In particular, security of current \pox schemes precludes any interrupts during the execution being proved.
We argue that lack of asynchronous capabilities (i.e., interrupts within PoX) can obscure PoX usefulness, as several applications require processing real-time and asynchronous events.
To bridge this gap, we propose, implement, and evaluate an \acrotextcapital (\acro).
\acro is secure under full software compromise, enables asynchronous \pox, and incurs less hardware overhead than prior work.
\end{abstract}

\maketitle

\renewcommand{\shortauthors}{Caulfield et al.}

\section{Introduction}\label{sec:intro}

Embedded (aka IoT or ``smart'') devices are increasingly popular worldwide and are becoming pervasive in all sorts of environments: from homes and offices to public spaces and industrial facilities. Not surprisingly, they are also increasingly targeted by exploits and malware. In particular, low-end micro-controller units 
(MCUs) are designed with strict cost, size, and energy limitations. Thus, it is hard to offer 
any concrete security guarantees for tasks performed by these MCUs, due to their lack of sophisticated 
security features, akin to those available to higher-end application processors, such as the ones used in smartphones or general-purpose controllers, e.g., Alexa or Nest. As these low-end MCUs become ubiquitous 
(especially in safety-critical settings), exploits that corrupt their integrity, e.g., to forge a sensed value or ``lie'' about having performed some expected actuation, become a significant threat.

Over the past decade, this problem was recognized and explored by the research 
community~\cite{RA2022survey}. Previous results considered potential unauthorized software modifications/compromises in low-end devices and proposed methods to remotely verify the binary currently installed in a low-end MCU: a well known security service referred to as \emph{Remote Attestation (\RA)}~\cite{smart,vrasedp,tytan,trustlite}.
While \RA can prove that a remote low-end MCU is currently installed with the proper software binary, it does not provide any proofs about the correct execution of this binary (or parts thereof, i.e., functions within the binary).
Therefore, recent work has focused on enhancing \RA architectures with the ability to prove the correct execution of the attested software~\cite{apex}, i.e., to generate Proofs of eXecution (\pox). The \pox capability, in turn, was shown to be a fundamental building block to provide additional guarantees, such as  
control- \& data-flow attestation~\cite{tinycfa,dialed}. We discuss both \RA and \pox in more detail in Section~\ref{sec:background}.

In addition, \pox can be used as a means to create sensors and actuators that ``can not lie'' even under the assumption that the MCU software implementing the sensing task may be compromised~\cite{apex}. This is because \pox enables generation of unforgeable proofs for the proper execution of software tasks, including their interaction with analog peripherals via General Purpose Input/Output (GPIO) interfaces. As these proofs also bind the execution to any generated outputs (e.g., sensed values), they serve as a cryptographic proof for the integrity of the sensing process as a whole, including peripheral configuration, acquisition, and processing of the raw data.

Despite these advances, thus far \pox has assumed that executables must run atomically and therefore do not process interrupts. As a consequence, tasks that require handling asynchronous inputs and events (e.g., the arrival of network packets or expiring timers), cannot benefit from \pox. On the other hand, most real embedded applications depend on interrupts to process asynchronous events due to real-time needs. Therefore, we pose a natural question:

\begin{center}
\vspace{0.2mm}
\noindent\textbf{\emph{Are secure proofs of execution attainable for executables that must process asynchronous and real-time events/inputs?}}
\vspace{0.2mm}
\end{center}

In this paper we set out to answer this question by designing  \acro: an \acrotextcapital.
At a high level, the proposed design introduces two new features to an existing \pox architecture (APEX~\cite{apex}), namely \emph{Ephemeral Immutability and Integrity for (1) the interrupt vector table (IVT); and (2) interrupt service routines (ISRs)}.
These features are achieved through the selective linking of relevant ISR binaries into specific protected (and attested) memory locations; attestation of IVT; and minimal (formally verified) additional hardware support.
As we discuss in the remainder of this paper, these features are sufficient to enable secure \pox that can also handle asynchronous events/inputs through the use of MCU interrupts. 
Our evaluation shows that \acro reduces the hardware overhead of existing \pox and incurs no additional run-time or storage/memory overhead. 

\section{Preliminaries}\label{sec:background}
%
%
\subsection{Scope of Low-End MCUs}\label{sec:scope}
This paper focuses on tiny CPS/IoT sensors and actuators, or hybrids thereof.
These are some of the smallest and weakest devices based on low-power single-core MCUs with small 
program and data memory (e.g., Atmel AVR ATMega, TI MSP430), with
$8$- and $16$-bit CPUs running at $1$-$16$MHz, with $\approx64$ KBytes of addressable memory.
SRAM is used as data memory, normally ranging between $4$ and $16$KBytes, while the rest of the
address space is available for program memory.  Such devices usually run software atop ``bare metal'', 
execute instructions in place  (physically from program memory), and lack memory management units (MMU) 
or privilege levels to support virtual memory or secure micro-kernels.

\subsection{Remote Attestation (\RA)}\label{sec:background_ra}
\RA allows a trusted verifier (\vrf) to detect unauthorized binary modifications (e.g., malware infections) 
on an untrusted remote device, called a prover (\dev) by remotely measuring the latter's software state.
Per Fig.~\ref{fig:timeline}, \RA is typically realized as a challenge-response protocol with the following steps:\\
%
	\noindent\textbf{1)-} \vrf sends an attestation request containing a challenge (\chal) to \dev. 
	This request might also contain a token derived from a secret that allows \dev to authenticate \vrf.
	
	\noindent\textbf{2)-} \dev receives the attestation request and computes an {\em authenticated integrity check} 
	over a predefined memory region (e.g., program memory) and \chal.
	
	\noindent\textbf{3)-} \dev returns the result to \vrf.
	
	\noindent\textbf{4)-} \vrf receives the result from \dev, and checks whether it corresponds to a valid memory state.
%
\begin{figure}[ht]
\vspace{-1em}
\centering
\scalebox{0.7}[0.7]{
	\fbox{
    \begin{tikzpicture}[node distance=1.5cm, >=stealth]
	\coordinate (BL)	at (0, 3);		\coordinate[left =4cm  of BL]	 (TL);
	\coordinate (Btcs)	at (0, 2.2);		\coordinate[left =4cm  of Btcs] (Ttcs);
	\coordinate (Btce)	at (0, .8);		\coordinate[left =4cm  of Btce] (Ttce);
	\coordinate (BR)	at (0, 0);		\coordinate[left =4cm  of BR]	 (TR);

	\node[above] at (BL) {\large Prover (\dev)};
	\node[above] at (TL) {\large Verifier (\vrf)};
	\coordinate (Chksum) at ($(Btcs)!0.5!(Btce)$);
	\node [right = .1cm, align=center] at (Chksum) {\small (2) Authenticated \\ \small Integrity Check \\ \small (e.g., MAC)};
	\coordinate (Verify)	at ($(Ttce)!0.5!(TR)$);
	\node [left =.5cm, align=center] at (Verify) {\small (4) Verify \\ \small Response};

	\draw[line width = .3cm, color=pink!50]	(BL) -- (BR);
	\draw[line width = .3cm, color=green!50!black] (TL) -- (TR);
	\coordinate (ReqStart) at ($(Ttcs)!0.1!(Btcs)$);
	\coordinate (ReqEnd) at ($(Ttcs)!0.9!(Btcs)$);
	\draw[thick, ->] (ReqStart) -- (ReqEnd) node [above=-.05cm, midway, sloped] {\small(1) Request};
	\coordinate (RepStart) at ($(Btce)!0.1!(Ttce)$);
	\coordinate (RepEnd) at ($(Btce)!0.9!(Ttce)$);
	\draw[thick, ->] (RepStart) -- (RepEnd) node [above=-.05cm, midway, sloped] {\small(3) Response};
\end{tikzpicture}
	}
}
\vspace{-1em}
	\caption{\small \RA interaction}
	\label{fig:timeline}
\vspace{-1.25em}
\end{figure}

The {\em authenticated integrity check} is usually realized as a Message Authentication Code (MAC) or a digital signature
over \dev memory. However, these cryptographic primitives require \dev to have a unique secret key (\attkey) 
either shared with \vrf (MAC-s), or for which \vrf knows the public key (signatures). This \attkey must reside in 
secure storage, and {\bf not} be accessible to any (potentially compromised) software running on \dev, except for trusted
attestation code itself. Since most \RA threat models assume a fully compromised software state on \dev, secure storage 
implies some level of hardware support.

\RA architectures fall into three categories depending on the level of hardware support: software-based, hardware-based, and 
hybrid. Security of software-based attestation~\cite{KeJa03, SPD+04, SLS+05, SLP08} relies on strong assumptions about 
precise timing and constant communication delays, which are unrealistic in the IoT/CPS ecosystem. Hardware-based 
methods~\cite{PFM+04, KKW+12, SWP08} rely on dedicated hardware components, e.g., TPMs~\cite{TPM}, Intel SGX~\cite{sgx}, 
or ARM TrustZone~\cite{trustzone}. However, the cost of such hardware is prohibitive for low-end MCU-s. 
Hybrid \RA~\cite{smart, tytan, vrasedp} aims to achieve security equivalent to hardware-based mechanisms, 
with low(er) hardware cost. It implements the authenticated integrity ensuring function in software, while relying on minimal 
hardware support to assure that this software implementation executes properly and securely.
\vspace{-.5em}
\subsection{Proofs of Execution (PoX)}\label{sec:background_pox}
PoX augments \RA's capability by proving to \vrf that: (1) the expected executable is stored in program memory, (2) this code has indeed executed, and (3) any claimed outputs were produced by its timely and authentic execution.

The first PoX architecture targeting low-end MCU-s was recently proposed in APEX~\cite{apex}.
APEX implements a hardware module controlling the value of a $1$-bit flag called $EXEC$, which cannot be written by any software. 
A value $EXEC=1$ indicates to \vrf that attested code \emph{must} have executed successfully, between the time when the 
challenge \chal was received from \vrf (recall the \RA protocol from Section~\ref{sec:background_ra}) and the time when the 
\RA measurement occurs (via authenticated integrity ensuring function). Similarly, when it receives an attestation reply with $EXEC=0$, 
\vrf can conclude that execution of said code did not occur, or that execution (or its output) was tampered with.
In APEX, the \RA measurement covers: {\bf(i)} the $EXEC$ flag; {\bf(ii)} the region where this execution's output is saved 
(output region -- $OR$); and {\bf(iii)} the executable itself (stored in the executable region -- $ER$). Thus, security of the 
underlying \RA architecture guarantees that the contents of these memory regions cannot be forged/spoofed to something 
different from their values at the time of the attestation computation. In turn, APEX considers a code to execute properly (and sets 
$EXEC=1$) if and only if:

	\noindent\textbf{1)-} Execution is \underline{atomic and uninterrupted}, from the first instruction (legal entry $ER_{MIN}$), to the last instruction (legal exit $ER_{MAX}$);
	
	\noindent\textbf{2)-} Neither the executable ($ER$), nor its outputs $OR$ are modified in 
	between the execution and subsequent \RA computation;
	
	\noindent\textbf{3)-} During execution, data-memory (including $OR$) is not modified, by other means except for the 
	ER execution, e.g., no modifications by other software or Direct Memory Access controllers.

From these conditions, $EXEC=1$ assures that memory contents (of $ER$ and $OR$) are consistent between $ER$'s 
code execution and subsequent \RA. It also assures that $ER$'s execution has integrity, e.g., it can not be corrupted by malicious interruptions that could alter its control-flow or its variables in data memory. $ER$ and $OR$ locations and sizes are configurable, allowing for PoX of arbitrary code 
and output sizes. APEX implementation builds atop the formally verified hybrid \RA architecture \vrased~\cite{vrasedp} 
and APEX hardware module is itself formally verified to adhere to a set of specifications. The conjunction of these properties are proven sufficient to imply a security notion (stated using a cryptographic 
security game~\cite{crypto_book}) for unforgeable proofs of execution. For brevity, we do not 
overview APEX proofs and refer the interested reader to~\cite{apex}.

As explained above, APEX mandates the absence of interrupts to guarantee that no untrusted interrupt sources and respective (potentially malicious) ISRs can interfere with the intended behavior of the executable (located in $ER$).
However, this also limits the types of executable behaviors for which \pox is possible. In particular, it prevents provable executions from leveraging any interrupts. 
In this work, we remedy this issue by enabling selected trusted ISRs, implementing intended asynchronous behavior, to be a part of provable executions without compromising \pox security.

\vspace{-0.5em}
\section{Application Examples}\label{sec:motivation}

Consider that \dev is a simple MCU implementing a syringe pump that allows remote physicians to monitor and deliver medication to patients over a network, e.g., as described in~\cite{islam2019design}.
Given the safety-critical and real-time nature of this application, it is paramount to verify that the execution of operations in \dev happens as expected, e.g., that \dev injects an \emph{accurate and timely} dosage to the patient;
otherwise, it may lead to under/overdose, affecting the patient's well-being.
Such an execution can be implemented as follows:

	\noindent\textbf{(1)-} Start injecting medication at a fixed rate;~\\
	\noindent\textbf{(2)-} Set up a timer interrupt according to the dosage to be injected;~\\
	\noindent\textbf{(3)-} Enter sleep/low-power mode;~\\
	\noindent\textbf{(4)-} Wake up once the timer expires and stop the injection.

To verify the correct execution of these steps, one may choose to implement \dev using an MCU equipped with a PoX architecture such as APEX (see Section~\ref{sec:background}).
However, doing so poses a challenge since this execution strictly depends on an asynchronous event (i.e., the expiring timer in step \textbf{(4)}), which in turn relies on a timer-based interrupt during execution.
Unfortunately, to ensure the integrity of execution, APEX prohibits interrupts during the execution of the software being proved. Therefore, it can not be used directly to provide security/safety guarantees in this example.


To enable APEX \pox for this application, one simple workaround is to modify \dev software as follows: instead of going to sleep and waking up based on the timer interrupt, 
\edit{the device uses the CPU to countdown, i.e., it \emph{busy-waits} for the expected period.} 

Doing so eliminates the need for a timer interrupt during execution and thus allows this application to benefit from APEX \pox.
Nonetheless, this workaround has important drawbacks.
\edit{First, it imposes an unnecessary power consumption by requiring the processor to actively wait and check for the critical event.}
This is a significant burden for battery-powered devices (e.g., portable insulin pumps).
Aside from the power consumption issue, in case of an emergency, the patient may choose to abort \dev execution, e.g., by pressing a physical ``cancel'' button or by sending a network command to ``abort''.
\edit{However, since the CPU is fully occupied and no interrupts are allowed during execution,}
\dev software has no way to detect/receive and process such asynchronous safety-critical command(s).
\edit{This illustrates why these approaches do not satisfy this application's safety-critical and real-time requirements.}

On the other hand, simply removing the \pox atomicity requirement from APEX opens the door for vulnerabilities. For example, after infecting \dev, malware may trigger an interrupt while the medication is being injected to increase the timer expiration value or, more generally, tamper with this execution by manipulating stored variables/parameters/data or its control-flow.

Aside from this example, it is not hard to find similar settings where the same real-time needs are applicable (e.g., industrial, automotive, etc). This general need motivates our work on the design of a secure \pox architecture that supports interrupts and thus can process asynchronous events/inputs.
\section{\acro Design}\label{sec:design}
\begin{figure}
  \centering
  \includegraphics[width=0.75\columnwidth]{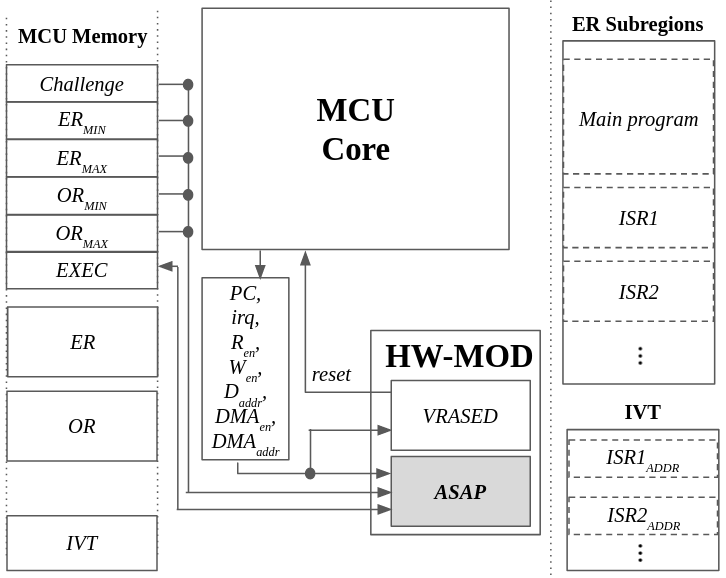}
  \vspace{-1em}
  \caption{\acro System Architecture}
  \vspace{-2em}
  \label{fig:system}
\end{figure}
%
%
MCUs process asynchronous events by either (1) \edit{{\it busy-waiting}}, i.e., actively checking in software; or \edit{(2)} via hardware interrupts. As discussed in Section~\ref{sec:motivation}, \edit{(1)} is, in many cases, not a viable approach.
To prevent abuse from software external to $ER$, APEX treats any interrupt as a violation. It sets $EXEC=0$ whenever the respective hardware signal (namely $irq$) is set, indicating an incoming interrupt during $ER$'s execution. 
We design \acro to improve \pox with the ability to:

	\noindent\textbf{(1)-} define which interrupts (and respective ISRs) are allowed and trusted as a part of $ER$ behavior.\\
	\noindent\textbf{(2)-} assure that the integrity of these allowed/trusted ISR(s) can be checked by \vrf as a part of the \pox result.\\
    \noindent\textbf{(3)-} guarantee that no other untrusted/unauthorized interrupt can occur during (or tamper with) $ER$ execution without \vrf detection.

Fig.~\ref{fig:system} presents \acro architecture at a high level.
\acro mandates that ISR binaries that are a part of $ER$ behavior be placed (linked) within the $ER$ memory region. As in Fig.~\ref{fig:system}, after compilation \& linking, $ER$ is composed of both the main program and all ISRs relevant to $ER$ execution.

With that, instead of checking for the value of the $irq$ to determine whether or not ``some interrupt has happened'', \acro can check the program counter ($PC$) value. If a trusted/authorized interrupt occurs, by construction $PC$ will remain inside $ER$ and \acro will keep $EXEC=1$ (valid \pox). If an untrusted/unauthorized interrupt occurs, $PC$ must leave $ER$. \acro will treat the latter as a violation and set $EXEC=0$.    
As the size of $ER$ is configurable (by setting the values of parameters $ER_{MIN}$ and $ER_{MAX}$), $ER$ size can be adjusted to fit the binaries of the main program + intended ISRs.
    

Furthermore, when an interrupt is triggered, the MCU fetches the address of an ISR from the IVT based on the hardware trigger source (e.g, GPIO, network/UART, timer, etc). Therefore, as a part of \pox, it is paramount to ensure that the contents of IVT (i.e., the addresses of functions that get called due to each type of interrupt) are also attested and that the content of IVT remains consistent from the time when $ER$ execution happens until when it is measured by the subsequent attestation
(recall the interplay between \pox and \RA discussed in Section~\ref{sec:background}).
\subsection{Adversary Model}\label{sec:adv}

We consider an adversary that controls \prv entire software state, including code and data. It can modify any writable memory and read any memory that is not explicitly protected by hardware-enforced access controls. Modifications to program memory can change instructions to modify the executable behavior whereas modifications to data memory can corrupt intermediate computation results or induce deviation from a program's intended control-flow. Finally, the adversary can attempt to change memory to program arbitrary interrupts before, during, or after a \pox.
\subsection{\acro Details}\label{sec:design_datails}

To enable processing of selected interrupts as a part of the \pox, \acro modifies APEX
atomicity requirements. We here go over these requirements, as well as \acro modifications in detail. APEX verified properties are specified in Linear Temporal Logic (LTL), which is particularly useful for specifying and verifying sequential systems. LTL extends common logic statements with temporal quantifiers. In addition to propositional connectives, such as conjunction ($\land$), disjunction ($\lor$), negation ($\neg$), and implication ($\rightarrow$), LTL includes temporal quantifiers, thus enabling sequential reasoning. In this paper, we consider the following two LTL quantifiers:
\begin{compactitem}
 \item \textbf{X}$\phi$ -- ne\underline{X}t $\phi$: holds if $\phi$ is true at the next system state.
 \item \textbf{G}$\phi$ -- \underline{G}lobally $\phi$: holds if for all future states $\phi$ is true.
\end{compactitem}

Atomicity and Uninterruptibility of $ER$ execution, as required by APEX, are formalized in LTL statements~\ref{eq:ephe_atom1},~\ref{eq:ephe_atom2}, and~\ref{eq:ephe_atom3}, per~\cite{apex}.
\begin{align}\label{eq:ephe_atom1}
\Small
\begin{split}
\text{\bf G}: \ \{(PC \in ER) \land \neg (\text{\bf X}(PC) \in ER) \rightarrow PC = ER_{MAX} \lor \neg \text{\bf X}(EXEC) \ \}
\end{split}
\end{align}
\begin{align}\label{eq:ephe_atom2}
\Small
\begin{split}
 & \text{\bf G}: \ \{\neg (PC \in ER) \land (\text{\bf X}(PC) \in ER) \rightarrow \text{\bf X}(PC) = ER_{MIN} \lor \neg \text{\bf X}(EXEC) \}
\end{split}
\end{align}
\begin{align}\label{eq:ephe_atom3}
\Small
\begin{split}
 & \text{\bf G}: \ \{(PC \in ER) \land irq \rightarrow \neg EXEC \}
\end{split}
\end{align}

The {\bf G} quantifier, surrounding all statements, requires them to hold at all times.
LTL~\ref{eq:ephe_atom1}
enforces that the only way for $ER$ execution to terminate without setting $EXEC=0$ is through its last instruction: $PC = ER_{MAX}$. This is specified by checking the relation between current and next $PC$ values. If the current $PC$ value is within $ER$ and next $PC$ value is outside $ER$, then either current $PC$ value is the address of $ER_{MAX}$, or $EXEC$ is set to $0$ in the next cycle.
Similarly, LTL~\ref{eq:ephe_atom2} uses {\bf X} quantifier to enforce that the only way for $PC$ to enter $ER$ is through the very first 
instruction: $ER_{MIN}$. This prevents $ER$ execution from starting at  some point in the middle of $ER$, thus 
ensuring that $ER$ always executes in its entirety. Finally, LTL~\ref{eq:ephe_atom3} enforces that $EXEC$ is 
set to zero if an interrupt happens during $ER$ execution, by checking the $irq$ signal.
To enable selected interrupts to be triggered securely, \acro removes LTL~\ref{eq:ephe_atom3} and adds two new requirements {\bf [AP1]} and {\bf [AP2]}.

\textbf{[AP1]: \textit{IVT Immutability \& Integrity}} - the memory region containing the IVT cannot be modified from the start of $ER$ execution until the end of attestation. This property ensures that the attestation result correctly portrays the addresses of all ISR(s) that could have been called and processed during $ER$ execution. Without this property, an adversary could modify IVT to cause an interrupt to jump to arbitrary locations within $ER$ leading to violations to $ER$ intended control-flow (and therefore $ER$ execution integrity). 
This new \acro property is formally specified in LTL~\ref{eq:ephe_immut}, based on signals that indicate a memory write to IVT by either the CPU or DMA. 
\begin{align}\label{eq:ephe_immut}
\Small
\begin{split}
\text{\bf G}: \{
[DMA_{en} \land (DMA_{addr} \in IVT)] \lor &\\ [W_{en} \land (D_{addr} \in IVT)] \rightarrow \neg EXEC\}
\end{split}
\end{align}
In LTL~\ref{eq:ephe_immut}, $W_{en}$ is a CPU signal that indicates that a CPU memory write is happening to the address  in the $D_{addr}$ signal. $DMA_{en}$ and $DMA_{addr}$ serve the same purpose for detecting writes by DMA to specific locations.
$EXEC$ is set to $0$ whenever there is a CPU write or DMA access to IVT.

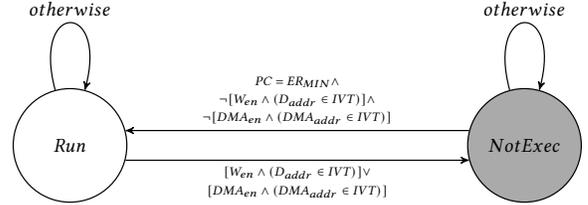
\begin{figure}
\begin{center}
\noindent\resizebox{0.9\columnwidth}{!}{%
	\begin{tikzpicture}[->,>=stealth',auto,node distance=8.0cm,semithick]
		\tikzstyle{every state}=[minimum size=2cm]
		\tikzstyle{every node}=[font=\large]

		\node[state] 		(A)					{$Run$};
		\node[state, fill={rgb:black,1;white,2}]         (B) [right of=A,align=center]	{$NotExec$};

		\path[->,every loop/.style={looseness=8}] 
			(A) edge [loop above] node {$otherwise$} (A)
			(B) edge [loop above] node {$otherwise$} (B);
  		
\draw[->] (A.345) -- node[rotate=0,below, align=center,auto=right] {\scriptsize \shortstack{$[W_{en} \land (D_{addr} \in IVT)] \lor$\\$[DMA_{en} \land (DMA_{addr} \in IVT)]$}} (B.195);
\draw[<-] (A.15) -- node[rotate=0,above] {\scriptsize \shortstack{$PC=ER_{MIN}\land$\\$\neg[W_{en} \land (D_{addr} \in IVT)] \land$\\$\neg[DMA_{en} \land (DMA_{addr} \in IVT)]$}} (B.165);
	\end{tikzpicture}
}
\vspace{-1em}
\caption{Verified FSM for LTL~\ref{eq:ephe_immut} for IVT Immutability.}
\vspace{-1em}
\label{fig:meta_fsm}
\end{center}
\end{figure}

Fig.~\ref{fig:meta_fsm} depicts a Verilog FSM implemented and verified to comply with LTL~\ref{eq:ephe_immut} (\textbf{[AP1]}).  The FSM has two states: $Run$ and $NotExec$. 
The FSM transitions to the $NotExec$ state and outputs $EXEC = 0$ whenever a violation happens, 
i.e., whenever $IVT$ is modified.
It transitions back to $Run$ when $ER$'s execution is restarted ($PC=ER_{MIN}$).

\textbf{[AP2]: \textit{ISR Immutability}} - trusted/authorized ISR binaries cannot be modified between the start of $ER$ execution until attestation is completed. This is required to ensure that the attestation result correctly reflects the behavior implemented by the ISRs as a part of $ER$. Without this property, the adversary could overwrite an authorized ISR arbitrarily, modifying its behavior without detection by \vrf. Since APEX already enforces $ER$ immutability between execution and attestation, \acro reuses this support by simply placing (linking) the trusted ISR binaries to be within $ER$.

\textit{\textbf{ASAP Security Argument:} Let $ER$ contain a program composed of a main task and its trusted ISR(s) which can be asynchronously executed due to their respective interrupt triggers. Once execution starts ($PC=ER_{MIN}$), APEX ensures that the \pox result will reflect $EXEC=1$ iff: (1) $ER$ is not modified until both its execution and subsequent attestation are over; (2) no external execution: $PC$ stays within $ER$ until it reaches $ER_{MAX}$ (LTL~\ref{eq:ephe_atom1}); and (3) $OR$ is not modified in between execution and attestation completion. 
Per {\bf [AP1]}, IVT is also immutable after $ER$ execution starts (otherwise \acro sets $EXEC=0$) and is attested.
Hence, the \pox result includes a report detailing which code section is executed due to each interrupt source in the system (i.e., the IVT configuration).
Finally, due to {\bf [AP2]}, ISRs relevant to $ER$ execution are all contained within $ER$, making them immutable and attested. Therefore, the \pox result allows \vrf to check that all IVT entries that point to an address within $ER$ correspond to the entry point of an intended/expected ISR binary. Additionally, any execution of an unauthorized/untrusted ISR requires jumping outside $ER$, which sets $EXEC=0$ (per LTL~\ref{eq:ephe_atom1}), resulting in an invalid \pox.}

\section{Implementation \& Evaluation}\label{sec:implementation}

We implemented \acro on OpenMSP430: an open-source design for the MSP430 architecture, which represents the targeted class of devices discussed in Section~\ref{sec:scope}.
\acro builds on top of APEX, which in turn relies on VRASED \RA architecture.
As shown in Fig.~\ref{fig:system}, \acro is implemented within the module labeled \hw. Its features are attained by small hardware modifications. \acro is publicly available at ~\cite{asapcode}.

To achieve {\bf [AP1] IVT Immutability}, the hardware is extended with the module shown in Fig.~\ref{fig:meta_fsm}, used to detect any writes to IVT. In OpenMSP430, IVT is stored in a fixed physical location, i.e., in the last 32-byte of addressable memory: from the base address 0xFFE0 to last address 0xFFFF.

\begin{figure}[t]
    \centering
	\subfigure[Linker Example]{\label{fig:linker}\includegraphics[width=0.46\columnwidth]{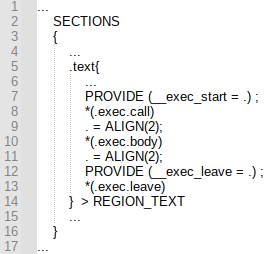}}
	\subfigure[Software Example]{\label{fig:software}\includegraphics[width=0.52\columnwidth]{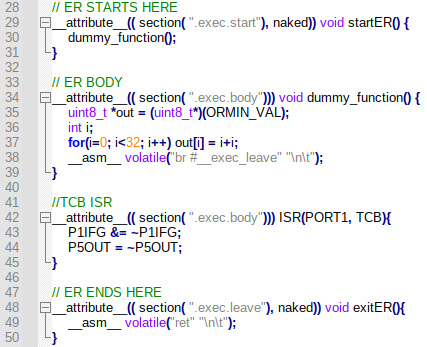}}
	\vspace{-1.5em}
	\caption{Software and Linking to achieve ISR immutability}\label{fig:software_all}
	\vspace{-1.5em}
\end{figure}

To selectively link trusted ISRs, as required by {\bf [AP2] ISR Immutability}, we implement $ER$ linking as in Fig.~\ref{fig:software_all}. In it, a sample ``dummy function'' executes a loop, and one ISR is implemented to write to GPIO PORT5 when any asynchronous signal is received from GPIO PORT1 (e.g., a button press). The header of these functions assigns them with the section label “exec.body”. This label in conjunction with a modified linker script for MSP430 allows for these functions to be placed inside $ER$ region. Labels “exec.start” and “exec.leave” are used to determine $ER$ entry point ($ER_{MIN}$) and exit point ($ER_{MAX}$). In this example, these functions are named \textit{startER()} and \textit{exitER()}. \textit{startER()} simply calls ``dummy function'' (i.e., the program's behavior) and \textit{exitER()} simply returns, i.e., concludes the provable execution. These functions have section labels “exec.start” and “exec.leave” so that they can be identified and placed at the beginning and end of $ER$ by the linker script shown in Fig.~\ref{fig:linker}.

\begin{figure}[t]
    \centering
	\subfigure[Authorized interrupt in ASAP]{\label{fig:asapval}\includegraphics[width=
	0.9\columnwidth]{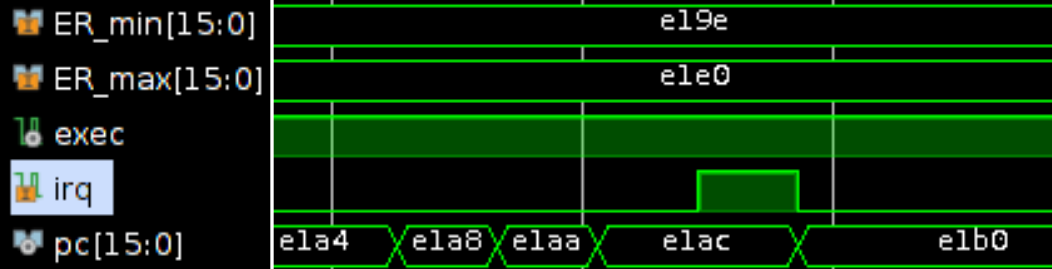}}
	\vspace{-1em}
	\subfigure[Unauthorized interrupt in ASAP]{\label{fig:asapinval}\includegraphics[width=
	0.9\columnwidth]{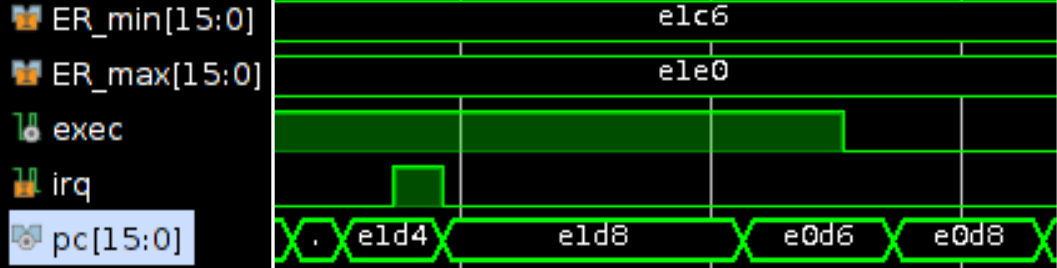}}
	\vspace{-1em}
	\subfigure[Any interrupt in APEX]{\label{fig:apexirq}\includegraphics[width=
	0.9\columnwidth]{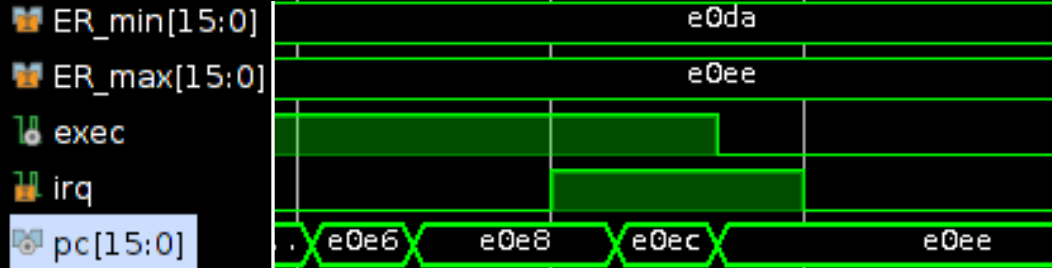}}
	\vspace{-.5em}
	\caption{Comparison: interrupt handling in \acro vs. APEX}\label{fig:exp}
	\vspace{-.5em}
\end{figure}

With this design in place, experiments were conducted to demonstrate the differences between APEX and \acro when processing interrupts during $ER$ execution. Fig.~\ref{fig:exp} shows simulation wave-forms for three cases including APEX and \acro. In each figure, the following signals are depicted over time: $ER_{MIN}$, $ER_{MAX}$, $EXEC$, $irq$ (interrupt request signal), and $PC$.

Fig.~\ref{fig:asapval} shows the behavior of \acro in the instance that a legitimate and authorized interrupt occurs (i.e., its corresponding ISR lies within ER) while executing ER. As shown, $ER$ is currently executing, as the $PC$ value is between $ER_{MIN}$ and $ER_{MAX}$. As the signal $irq$ is set, indicating an interrupt, $PC$ jumps from the main program at $0xE1AC$ to the ISR first instruction at $0xE1B0$. Since the destination is still within the range of the $ER$, the $EXEC$ signal is unaffected and remains 1. As a result, a subsequent attestation would convey to \vrf that execution was successful and untampered with. 
Fig.~\ref{fig:asapinval} shows the behavior of \acro under the influence of an external interrupt that has not been authorized to be a part of $ER$ behavior and therefore not linked within $ER$. 
In this scenario, initially $PC$ is also within $ER$. However, once the external interrupt is handled, $PC$ jumps to the ISR located outside of $ER$ (at $0xE0D6$). In accordance with LTLs~\ref{eq:ephe_atom1}~and~\ref{eq:ephe_atom2}, $EXEC$ is set to $0$. In APEX, shown in Fig.~\ref{fig:apexirq}, any $irq$ causes $EXEC=0$, regardless of the $PC$ value or whether or not the ISR is located within $ER$ and a part of the executable behavior.
This illustrates \acro ability to separate trusted and untrusted interrupts and handle trusted interrupts while keeping \pox secure against untrusted ones.

To demonstrate \acro practicality, we synthesized and implemented its RTL design on an Artix-7 FPGA (Basys3 prototyping board). We note that a hardware design that is synthesizable on FPGA can also be used to manufacture an Application-Specific-Integrated-Circuit (ASIC) for large-scale usage. Below we report on \acro costs based on this prototype.

\begin{figure}[t]
    \centering
	\subfigure[Total extra Look-Up Tables (LUTs)]{\label{fig:lutcomp}\includegraphics[width=0.49\columnwidth]{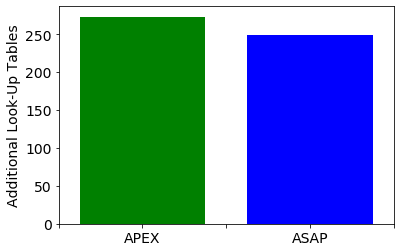}}
	\subfigure[Total extra registers]{\label{fig:reg}\includegraphics[width=0.49\columnwidth]{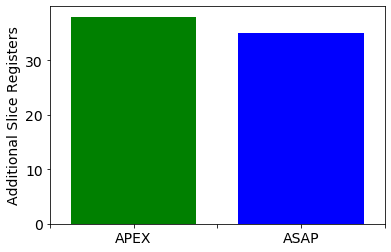}}
	\vspace{-1em}
	\caption{Overhead comparison between APEX and \acro}
	\vspace{-1.5em}
	\label{fig:overhead}
\end{figure}

\textbf{Hardware and Memory Overhead. } 
To evaluate \acro hardware overhead, we compare it to APEX in Fig.~\ref{fig:overhead}.
Similar to related work~\cite{vrasedp,dessouky2017fat,dessouky2018litehax,apex,rata}, we 
consider the hardware overhead in terms of additional Look-up Tables (LUTs) and registers. The increase in LUTs is an estimate of the additional chip cost and size required for combinatorial logic, while the number of extra registers indicates additional states required by the sequential logic in \acro FSMs.
Fig.~\ref{fig:overhead} shows that \textit{\acro} utilizes 24 less LUTs and 3 less registers than APEX.
As \textbf{[AP2]} reuses existent $ER$ protection to ensure immutability of ISRs, it incurs no additional hardware overhead. Additionally, APEX requires monitoring the $irq$ signal, which is propagated into several sub-modules to enforce LTL~\ref{eq:ephe_atom3}. Because this is no longer required in \acro, there is a reduction in the register and LUT utilization, despite the need for an additional 2-state FSM to enforce {\bf [AP1]}.

\textbf{Runtime Overhead.} Neither \acro nor APEX incur additional execution time for the tasks being proved, as no instrumentation or additional instructions are required. This is because relevant runtime security properties and control of the $EXEC$ flag are implemented by hardware that runs in parallel with the CPU (as in Fig.~\ref{fig:system}). Linking in \textbf{[AP2]} is static and done at compilation time.

\textbf{Verification Cost.} 
We verified \acro on a 64-bit Ubuntu 18.04 machine with an Intel i7 3.6GHz CPU using NuSMV~\cite{nusmv} model-checker to show it adheres to the new property and still maintains all other guarantees required by APEX. 
\acro verification takes $\approx$150s for a total of 21 LTL properties and requires 96MB of RAM. \acro verified implementation totals 2155 lines in the Verilog HDL.
%
%
\vspace{-2.5em}
\section{Related Work}\label{sec:rw}
\textbf{\RA \& Interrupts.} One important security property of \RA (see Section~\ref{sec:background}) is \emph{temporal consistency}, i.e., guaranteeing that an \RA result always reflects an instantaneous snapshot of \prv attested memory. Lack thereof allows malware 
to escape detection by copying and/or erasing itself during \RA.
Temporal consistency is usually achieved by enforcing atomic (uninterruptible) \RA execution.
However, since \RA is often used in safety-critical and/or real-time settings~\cite{carpent2018reconciling}, the atomicity requirement might interfere 
with the MCU applications. To address this issue, SMARM~\cite{smarm} allows \RA to be interruptible 
by using probabilistic malware detection. Meanwhile, ERASMUS~\cite{erasmus} and SeED~\cite{ibrahim2017seed} 
are based on periodic self-measurements in order to detect transient malware that infects \prv and leaves before the next \RA instance. RATA~\cite{rata} actively monitors writes to program memory to detect such attacks.
We note that these efforts should not be confused with interruptable \pox. \RA by itself does not provide any runtime guarantees (see Section~\ref{sec:background}) but rather serves as a building block for more expressive proofs such as \pox.

\textbf{Proof of Execution (PoX).} Prior work has focused on providing PoX in high-end devices.
Flicker~\cite{flicker} leverages TPM and hardware extensions to implement a PoX functionality on Intel and AMD computers. Sanctum~\cite{sanctum} implements PoX in Intel-SGX-like devices by instrumenting enclaved code to output information about its own execution to a remote \vrf. 
Both of these approaches rely on complex hardware that is unavailable in low-end embedded systems.
Thus far, APEX~\cite{apex} is the only PoX architecture designed for simple MCUs. However, in order to successfully produce an unforgeable proof, APEX requires execution to run without interruptions, precluding its usefulness on interrupt-based applications.

\textbf{Control-Flow \& Data-Flow Attestation (CFA/DFA).}
The goal of CFA~\cite{cflat,dessouky2017fat}
is to measure the exact control flow path taken during execution on \dev. The result of this measurement can convey to \vrf the order in which instructions of some specific binary have executed on \dev.
DFA~\cite{oat,dessouky2018litehax} augments CFA with the ability to detect data-only attacks. 
Recent work~\cite{tinycfa, dialed}, built on top of APEX to obtain CFA and DFA in low-end MCUs. As such, they are also limited to APEX's uninterruptability requirement and can not process asynchronous inputs during execution/CFA/DFA. We believe that our work addresses this limitation.

\vspace{-0.5em}
\section{Conclusion}\label{sec:conclusion}

This work is motivated by much needed integrity assurance for execution in safety-critical edge devices,
which are often implemented with low-power and low-cost MCUs.
Existing mechanisms for producing unforgeable proofs of execution in low-end MCUs require executables not to handle interrupts, precluding their uses in many real-time/mission-critical settings.
To address this issue, we proposed \acro: the first architecture able to generate \pox for software that implements interrupts within its behavior.
\acro extends the original APEX \pox architecture to securely convey information about the ephemeral immutability and integrity of the IVT and relevant ISRs. We show that these two properties are sufficient to realize \acro securely and that they can be obtained through the appropriate linking of ISRs into protected memory and minimal additional verified hardware support, on top of that already provided by APEX.
Our experimental results show \acro feasibility and affordability, even on some of the lowest-end MCUs. 



\bibliographystyle{IEEEtranS}

	\bibliography{IEEEabrv,references}

\begin{thebibliography}{10}
\providecommand{\url}[1]{#1}
\csname url@samestyle\endcsname
\providecommand{\newblock}{\relax}
\providecommand{\bibinfo}[2]{#2}
\providecommand{\BIBentrySTDinterwordspacing}{\spaceskip=0pt\relax}
\providecommand{\BIBentryALTinterwordstretchfactor}{4}
\providecommand{\BIBentryALTinterwordspacing}{\spaceskip=\fontdimen2\font plus
\BIBentryALTinterwordstretchfactor\fontdimen3\font minus
  \fontdimen4\font\relax}
\providecommand{\BIBforeignlanguage}[2]{{%
\expandafter\ifx\csname l@#1\endcsname\relax
\typeout{** WARNING: IEEEtranS.bst: No hyphenation pattern has been}%
\typeout{** loaded for the language `#1'. Using the pattern for}%
\typeout{** the default language instead.}%
\else
\language=\csname l@#1\endcsname
\fi
#2}}
\providecommand{\BIBdecl}{\relax}
\BIBdecl

\bibitem{asapcode}
``{ASAP} source code,'' \url{https://github.com/RIT-CHAOS-SEC/ASAP}.

\bibitem{cflat}
T.~Abera~et al., ``C-flat: Control-flow attestation for embedded systems
  software,'' in \emph{ACM CCS}, 2016.

\bibitem{trustzone}
\BIBentryALTinterwordspacing
{Arm Ltd.}, ``Arm {TrustZone},'' 2018. [Online]. Available:
  \url{https://www.arm.com/products/security-on-arm/trustzone}
\BIBentrySTDinterwordspacing

\bibitem{tytan}
F.~Brasser~et al., ``Tytan: Tiny trust anchor for tiny devices,'' in
  \emph{DAC}.\hskip 1em plus 0.5em minus 0.4em\relax ACM, 2015.

\bibitem{erasmus}
X.~Carpent~et al., ``{ERASMUS}: Efficient remote attestation via
  self-measurement for unattended settings,'' in \emph{DATE}, 2018.

\bibitem{carpent2018reconciling}
------, ``Reconciling remote attestation and safety-critical operation on
  simple iot devices,'' in \emph{DAC}, 2018.

\bibitem{smarm}
------, ``Remote attestation of iot devices via {SMARM}: Shuffled measurements
  against roving malware,'' in \emph{IEEE HOST}, 2018.

\bibitem{nusmv}
A.~Cimatti~et al., ``Nusmv 2: An opensource tool for symbolic model checking,''
  in \emph{CAV}, 2002.

\bibitem{sanctum}
V.~Costan~et al., ``Sanctum: Minimal hardware extensions for strong software
  isolation,'' in \emph{$\{$USENIX$\}$ Security}, 2016.

\bibitem{vrasedp}
I.~De~Oliveira Nunes~et al., ``{VRASED}: A verified hardware/software co-design
  for remote attestation,'' \emph{USENIX Security}, 2019.

\bibitem{apex}
------, ``{APEX}: A verified architecture for proofs of execution on remote
  devices under full software compromise,'' in \emph{{USENIX} Security}, 2020.

\bibitem{dialed}
------, ``Dialed: Data integrity attestation for low-end embedded devices,''
  \emph{DAC}, 2021.

\bibitem{rata}
------, ``On the toctou problem in remote attestation,'' in \emph{ACM CCS},
  2021.

\bibitem{tinycfa}
------, ``Tiny-cfa: A minimalistic approach for control flow attestation using
  verified proofs of execution.'' in \emph{DATE}, 2021.

\bibitem{dessouky2017fat}
G.~Dessouky~et al., ``Lo-fat: Low-overhead control flow attestation in
  hardware,'' in \emph{DAC}, 2017.

\bibitem{dessouky2018litehax}
------, ``Litehax: Lightweight hardware-assisted attestation of program
  execution,'' in \emph{ICCAD}, 2018.

\bibitem{smart}
K.~Eldefrawy~et al., ``Smart: Secure and minimal architecture for (establishing
  dynamic) root of trust,'' in \emph{NDSS}, 2012.

\bibitem{ibrahim2017seed}
A.~Ibrahim~et al., ``{SeED}: secure non-interactive attestation for embedded
  devices,'' in \emph{ACM WiSec}, 2017.

\bibitem{sgx}
\BIBentryALTinterwordspacing
Intel, ``{I}ntel {Software Guard Extensions} ({Intel} {SGX}).'' [Online].
  Available: \url{https://software.intel.com/en-us/sgx}
\BIBentrySTDinterwordspacing

\bibitem{islam2019design}
M.~R. Islam~et al., ``Design and implementation of low cost smart syringe pump
  for telemedicine and healthcare,'' in \emph{ICREST}, 2019.

\bibitem{crypto_book}
J.~Katz and Y.~Lindell, \emph{Introduction to modern cryptography}.\hskip 1em
  plus 0.5em minus 0.4em\relax CRC press, 2014.

\bibitem{KeJa03}
R.~Kennell~et al., ``Establishing the genuinity of remote computer systems,''
  in \emph{USENIX Security}, 2003.

\bibitem{trustlite}
P.~Koeberl~et al., ``{TrustLite}: A security architecture for tiny embedded
  devices,'' in \emph{EuroSys}, 2014.

\bibitem{KKW+12}
X.~Kovah~et al., ``New results for timing-based attestation,'' in \emph{IEEE
  S\&P '12}, 2012.

\bibitem{RA2022survey}
B.~Kuang~et al., ``A survey of remote attestation in internet of things:
  Attacks, countermeasures, and prospects,'' \emph{Computers \& Security},
  2022.

\bibitem{flicker}
J.~McCune~et al., ``Flicker: An execution infrastructure for tcb
  minimization,'' in \emph{EuroSys}, 2008.

\bibitem{PFM+04}
J.~Petroni~et al., ``Copilot --- {A} coprocessor-based kernel runtime integrity
  monitor,'' in \emph{USENIX Security}, 2004.

\bibitem{SWP08}
D.~Schellekens~et al., ``Remote attestation on legacy operating systems with
  trusted platform modules,'' \emph{Science of Comp. Programming}, 2008.

\bibitem{SPD+04}
A.~Seshadri~et al., ``{SWATT}: {S}oftware-based attestation for embedded
  devices,'' in \emph{IEEE S\&P '04}, 2004.

\bibitem{SLS+05}
------, ``Pioneer: verifying code integrity and enforcing untampered code
  execution on legacy systems,'' in \emph{ACM SOSP}, 2005.

\bibitem{SLP08}
------, ``{SAKE:} software attestation for key establishment in sensor
  networks,'' in \emph{DCOSS}, 2008.

\bibitem{oat}
Z.~Sun~et al., ``Oat: Attesting operation integrity of embedded devices,'' in
  \emph{IEEE S\&P}, 2020.

\bibitem{TPM}
\BIBentryALTinterwordspacing
{TCG}, ``Trusted platform module (tpm),'' 2017. [Online]. Available:
  \url{http://www.trustedcomputinggroup.org/work-groups/trusted-platform-module/}
\BIBentrySTDinterwordspacing

\end{thebibliography}
\end{document}